\documentclass[twocolumn,prb,showpacs,multicol,amsmath,amssymb]{revtex4}
\usepackage[dvips]{graphicx}
\usepackage{sidecap}
\usepackage[usenames]{color}
\newcommand{\be}{\begin{equation}}
\newcommand{\ee}{\end{equation}}

\newcommand{\bea}{\begin{eqnarray}}
\newcommand{\eea}{\end{eqnarray}}
\newcommand{\bd}{\begin{displaymath}}
\newcommand{\ed}{\end{displaymath}}
\newcommand{\ba}{\begin{array}}
\newcommand{\ea}{\end{array}}
\newcommand{\bi}{\begin{itemize}}
\newcommand{\ei}{\end{itemize}}
\newcommand{\bc}{\begin{center}}
\newcommand{\ec}{\end{center}}
\newcommand{\bfl}{\begin{flushleft}}
\newcommand{\efl}{\end{flushleft}}
\newcommand{\bfr}{\begin{flushright}}
\newcommand{\efr}{\end{flushright}}

\def\6{\partial}

\def\={\!\!\!&=&\!\!\!}
\def\+{\!\!\!&&\!\!\!+~}
\def\-{\!\!\!&&\!\!\!-~}

\begin{document}
 \date{\today}
 \title{Quasiparticle interference in iron-based superconductors}

\author{A. Akbari$^1$, J. Knolle$^2$, I. Eremin$^1$, and R. Moessner$^2$}
\affiliation {$^1$Institut f\"ur Theoretische Physik III, Ruhr-Universit\"at Bochum, D-44801 Bochum, Germany}
 \affiliation{
 $^{2}$Max Planck Institute for the Physcis of Complex Systems, D-01187 Dresden, Germany}

 \begin{abstract}
 We systematically calculate quasiparticle interference (QPI) signatures for the whole phase
 diagram of iron-based superconductors. Impurities inherent in the sample together with ordered phases
 lead to distinct features in the QPI images that are believed to be measured in spectroscopic imaging-scanning
 tunneling microscopy (SI-STM). In the spin-density wave phase the rotational symmetry of the electronic
 structure is broken, signatures of which are also seen in the coexistence regime with both superconducting
 and magnetic order. In the superconducting regime we show how the different scattering behavior for magnetic
 and non-magnetic impurities allows to verify the $s^{+-}$ symmetry of the order parameter.
 The effect of possible gap minima or nodes is discussed.
 \end{abstract}

 \maketitle

\section{Introduction}
\begin{figure}[t]
\centering
 \includegraphics[width=.70\linewidth]{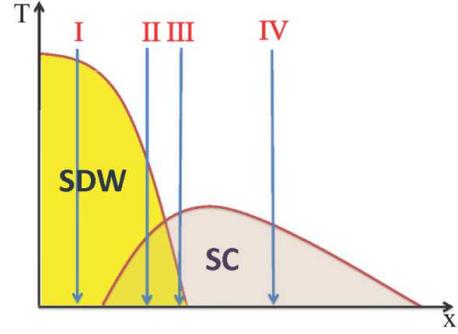}
\caption{(color online) Schematic phase diagram for the iron-based superconductors
[Ba(Fe$_{1-x}$Co$_x$)$_2$As$_2$ and  SmFeAs(O$_{1-x}$F$_x$)]
exhibiting a regime of coexistence  of  spin density wave (SDW) and superconducting (SC) orders\cite{local,bulk}.
Regimes (I)-(IV) refers to
the various parts of the phase diagram.}
\label{Phase}
\end{figure}


The relation between unconventional superconductivity
and magnetism is one of the most interesting topics in
condensed-matter physics.  In iron-based superconductors\cite{kamihara08} this question is of particular interest
because in some of these systems like Ba(Fe$_{1-x}$Co$_x$)$_2$As$_2$
and  SmFeAs(O$_{1-x}$F$_x$) superconductivity (SC)
and {\it metallic} antiferromagnetism (AFM) coexist homogeneously\cite{local,bulk}. This coexistence is characterized by a competition of the two ordered
states which results in a
dramatic suppression of the magnetization below the SC transition temperature, $T_c$,
to the extent that reentrance of the non-magnetically
ordered superconducting phase sets in at low temperatures\cite{fernandes}.

A number of groups have argued\cite{theory1,theory2,theory2a,theory3,theory4} that in the ferropnictides (FP), the
same electrons contribute to the superconducting condensate
and to the formation of the ordered magnetic moment.
Furthermore, it was shown\cite{fernandes,theory3,theory4} that a coexistence of the two phases
occurs only for a large set of parameters when the superconducting order
is of the $s^{+-}$-type, {\it i.e.} the superconducting gap changes sign between hole
and electron Fermi surface pockets which are located around the $\Gamma$ and $(\pm \pi,0)$,
($0,\pm \pi$) points of the Brillouin Zone (BZ) based on 1 Fe ion per unit cell (so-called {\it unfolded} BZ).
Therefore, a large coexistence region in the phase diagram of FPs indirectly supports the $s^{+-}$-wave
symmetry of the SC order in these systems.

The AFM (or SDW) phase itself shows a number of interesting properties which still have not been
completely understood. Given the electronic structure of FPs, consisting of 2 hole and 2 electron Fermi
surface pockets located around the $\Gamma$ and $(\pm \pi,0)$, ($0,\pm \pi$) points of the BZ\cite{LDA}, it is
natural to assume that
AFM order emerges, at least partly,  due to near-nesting between
the dispersions of holes and
electrons~\cite{theory1,theory2,lp,d_h_lee,timm1, timm2, timm3,dagotto,Korshunov2008,mj,honerkamp}. Within this
scenario the selected magnetic order is either with momentum ${\bf Q}_1 = (\pi,0)$ or ${\bf Q}_2= (0,\pi)$
[Ref. \onlinecite{theory2a}] which corresponds to ferromagnetic order along one and antiferromagnetic order
along the other direction. The collective spin excitations in the ordered state reveal pronounced anisotropy
along $x$ and $y$ directions which is a consequence of the fact that  only one of the two
elliptic pockets is involved in the formation of the AFM state\cite{knolle1}. Similarly, the electronic structure
in the AFM ordered state is anisotropic and the quasiparticle interference (QPI) introduced by scalar nonmagnetic,
as well
as magnetic, impurities gives rise to a pronounced quasi-one dimensional periodic modulation
of the real space local density of states which, as we argued in Ref.\onlinecite{Knolle2010}, is in agreement
with imaging-scanning tunneling microscopy (SI-STM) experiments in Ca(Fe$_{1-x}$Co$_x$)$_2$As$_2$\cite{chuang}.

Overall, QPI has become a powerful experimental tool for elucidating the nature of the many-body states
in novel superconductors. In the
presence of impurities, elastic scattering mixes two quasiparticle eigenstates
with momenta {\bf k}$_1$ and {\bf k}$_2$ on a
contour of constant energy. The resulting interference at wavevector ${\bf q} = {\bf k}_2 - {\bf k}_1$
reveals a modulation of the local density of
states (LDOS). The
interference pattern in momentum space can
be visualized by means of the SI-STM\cite{mkelroy}. In layered cuprates the analysis of the QPI
has provided details of the band structure,
the nature of the superconducting gap, or other competing
orders\cite{cuprates,cuprates1,cuprates2,cuprates3}.

In the superconducting state of iron-based superconductors a recent SI-STM experiment claimed to unambiguously
identify the order parameter symmetry to be of the $s^{+-}$ character\cite{hanaguri2010}.
Theoretical analyzes performed previously have also shown a pronounced difference in  interference patterns
characteristic of a simple $s$-wave symmetry and an extended $s^{+-}$-wave symmetry
[see Refs.\onlinecite{Zhang08,pereg,wang}].
In particular, in the presence of vortices an additional source of scattering either suppresses or enhances
intra- and inter-pocket scattering depending on the symmetry of the superconducting gap\cite{pereg}, a
feature found also in the experimental data\cite{hanaguri2010}. At the same time, the experimental results have been
put into question and argued instead to arise from the Bragg scattering and not due to QPI\cite{mazincomm}.

In what follows we address signatures of the different orders on the electronic structure as seen
in the distinctive features probed by QPI .
In particular, we extend our previous analysis of the QPI in the AFM state\cite{Knolle2010} to the
entire phase diagram of the FPs assuming that the AFM and SC arise from the same quasiparticles.
We present the results for QPI in the coexistence regime of AFM$+$SC
with $s$- and extended $s$-wave ($s^{+-}$) symmetry as well as for the pure superconducting state.
We further investigate the role of higher harmonics (non-trivial angular dependence along the Fermi surface)
in the extended $s-$wave case and, in particular,
we address the question whether a nodeless or nodal $s^{+-}$-wave symmetry can be identified within
SI-STM. The aim of our analysis is to find subtle features of the various many-body states in
iron-based superconductors that can be probed by STM.

The paper is organized as follows.
The theoretical calculations to obtain the local density of state (LDOS) are based on a four band model, and
they are explained in Sect. II.
Based on our model we numerically analyze the QPI signatures in the various phases in Sect. III.
We conclude the paper by a summary (Sect. IV).

\section{Theoretical Model}

We employ an effective mean-field four band model with two circular hole
pockets centered around  the $(0,0)$ point of the unfolded BZ  ($\alpha$-bands) and two elliptic electron
pockets centered at $(\pm \pi,0)$ and $(0,\pm \pi)$ points of the BZ, respectively ($\beta$-bands)\cite{knolle1}:
\begin{eqnarray}
\label{eqH}
H_c  & = &  \sum_{\mathbf{k}, \sigma,
i=\alpha_1,\alpha_2,\beta_1,\beta_2} \varepsilon^{i}_{\mathbf{k}} c_{i \mathbf{k}  \sigma}^\dag c_{i \mathbf{k} \sigma} +
\nonumber\\
&& \sum_{\mathbf{k}, \sigma,
i=\alpha_1,\alpha_2,\beta_1,\beta_2}
\left[\Delta^{i}_{\mathbf{k}} c_{i \mathbf{k}  \sigma}^\dag c_{i -\mathbf{k} -\sigma}^\dag+H.c.\right]+
\nonumber\\
&&  \sum_{{\bf k} \sigma}
W \sigma
\left[ c^\dag_{\alpha_1 {\bf  k} \sigma} c_{\beta_1 {\bf k}+{\bf Q}_1 \sigma}+ H.c.\right]
\end{eqnarray}
Here, we set the electronic dispersions to
$\varepsilon^{\alpha_i}_{\mathbf{k}} =\varepsilon_0 + t_\alpha\left( \cos k_x +\cos k_y \right) -\mu_i$ and
$\varepsilon^{\beta_1}_{\mathbf{k}}= \varepsilon_0 +
t_\beta\left( \left[1+\epsilon \right]\cos(k_x+\pi)+\left[ 1-\epsilon \right]\cos(k_y)\right) -\mu_1$,
$\varepsilon^{\beta_2}_{\mathbf{k}}= \varepsilon_0 +
t_\beta\left( \left[1-\epsilon \right]\cos(k_x)+\left[ 1+\epsilon \right]\cos(k_y+\pi) \right)-\mu_1$.
$\epsilon$ accounts for the ellipticity of the electron pockets.
Following our previous analysis of the spin excitations and QPI in the magnetically ordered state,
we use Fermi velocities and size of the Fermi pockets based on Refs.~\cite{LDA},
namely $t_\alpha=0.85 eV$, $t_\beta=-0.68 eV$, $\mu_{\alpha_1}=1.54 eV$, $\varepsilon_0 = 35$meV,
$\mu_{\alpha_2}=1.44 eV$, $\mu_1=-1.23 eV$, and $\epsilon=0.5$.
For these values, the Fermi velocities are $0.5eV a$ for the $\alpha_1$-band, where $a$ is the lattice
spacing, and $v_x=0.27 eV a$ and $v_y=0.49 eV a$ along  $x$- and $y$-directions for
the $\beta_1$-band, and vice versa for $\beta_2$. We use $a_x = a_y =a=1$.

Here, $\Delta^{i}_{\mathbf{k}}$ is the superconducting gap for band $i$. Symmetry and structure of the
superconducting gap in $Fe-$based superconductors have been subject of numerous experimental and theoretical
papers in recent years~\cite{kamihara08,mazin,scalapino,chubukov,wang,lp,d_h_lee}. There is a growing consensus
that the gap has an extended $s-$wave symmetry -- it belongs to a symmetric $A_{1g}$ representation of the $D_{4h}$
symmetry group of a square lattice. Gap values along electron and hole Fermi surfaces (FS) are of
opposite signs.

A more subtle issue  is whether the gap has nodes. This is not a symmetry issue as adding higher harmonics to
the extended $s-$wave gap yields stronger momentum dependence of the latter on the FS.
Quite generally one can write the superconducting gap in the form
\begin{displaymath}
 \Delta_{\mathbf{k}}=\Delta\left[\cos k_x \cos k_y \right] + \Delta^\prime[\cos(k_x)+\cos(k_y)]/2.
\end{displaymath}
For $\Delta^{\prime}=0$ the superconducting gap can be roughly approximated by a constant on the hole and electron
FS pockets but with opposite signs, $\Delta^e=-\Delta^h \simeq - \Delta$.
Increasing $\Delta^{\prime}$ does not significantly change the momentum
dependence of the gap on the FS of the hole pockets which can be approximated as
$\Delta^{\alpha_i}=\Delta^{h} $ while it introduces substantial angular dependence of the
superconducting gap on the electron pockets in the form  $\Delta^{\beta_i} (\psi) = \Delta^e (1 \pm b \cos 2 \phi)$.
Here, $\phi$ is the angle measured from the line connecting the two electron FSs, {\it i.e.} the line between $(\pi,0)$ and $(0,\pi)$ points of the first BZ and $b\simeq 2\Delta^\prime/ \Delta $.
Such $\Delta^{\beta_i} (\phi)$ has no nodes if $b<1$ but ``accidental'' nodes do appear when $b >1$.
Their positions are determined  by $\cos 2\phi = 1/b$ and the latter shift
continuously with $b$, so that the node's location is not selected by any symmetry.

FPs are multi-orbital systems and the orbital nature of the quasiparticle states introduces a sizable variation
of the interaction along the Fermi surface. This manifests itself as a $\cos 2\phi$ component of the
interaction. In particular, $b$ grows upon inclusion of the intra-band Coulomb repulsion into the gap equation
because this term couples  to the gap averaged
over the FS and hence reduces angle-independent gap components but does not affect $\cos 2\phi$
components~(Ref.\onlinecite{chubukov}).  As a consequence,
$b$ becomes progressively larger as the system moves further away from the SDW phase and the effect of
intra-band repulsion grows. In other words overdoped FPs are more likely to have nodes in the gap.

We introduce the experimentally observed $(\pi,0)$ SDW order parameter, within a
standard mean-field approximation: ${\vec W} \propto \sum_{\bf p} \langle c^\dag_{\alpha_1\mathbf{p}\delta}
c_{\beta_1 \mathbf{p+Q_1} \gamma} \vec{\sigma}_{\delta \gamma} \rangle$. In this state, one of the $\alpha$
fermions couples with only one band of $\beta$ fermions, leaving the other hole and electron bands --
and hence their electron and hole FSs -- unaffected by the SDW. Without loss of generality we
direct $\vec{W}$ along the $z$- quantization axis.

The actual QPI which is believed to be measured in SI-STM\cite{mkelroy} arises from quasiparticle
scattering by
perturbations internal to the sample such as non-magnetic or magnetic impurities. We perform a
standard analysis of such processes based on a T-matrix description\cite{vekhter}. In particular,  we introduce an
impurity term in the Hamiltonian
\bea
{\cal H}_{imp}=\sum\limits_{{\bf k} {\bf  k}^\prime i i^\prime\sigma\sigma^\prime}
\left(
V^{i i^\prime}_{ {\bf  k}{\bf  k}^\prime}
\delta_{\sigma\sigma^\prime}+ J^{i i^\prime}_{\sigma\sigma^\prime }{\bf S}\cdot {\bf \sigma}_{\sigma\sigma^\prime}
\right)
c^\dag_{i{\bf  k}\sigma} c_{i^{\prime}{\bf  k}^\prime\sigma^\prime}
\eea
where $V^{i i^{\prime}}_{ {\bf  k}{\bf  k}^\prime}$ and  $J^{i i^\prime}_{\sigma\sigma^\prime }$
represent the non-magnetic and the magnetic  point-like
scattering between the electrons in the bands $i$, and $i^\prime$ respectively.
In the following we set the quantization axis of the magnetic impurity along the $z$-direction.
 Following Ref. \onlinecite{vekhter}, we define the new Nambu spinor as
$
\hat{\psi}_{{\bf k}}^{\dagger}=(
c_{\alpha_2 {\bf k}\uparrow }^{\dagger},
c_{\alpha_1{\bf k}\uparrow}^{\dagger},
c_{\beta_1{\bf k}\uparrow  }^{\dagger},
c_{\beta_2 {\bf k}\uparrow }^{\dagger},
 c_{\alpha_2 -{\bf k}\downarrow },
 c_{\alpha_1 -{\bf k}\downarrow  }
 ,c_{\beta_1 -{\bf k}\downarrow  },
  c_{\beta_2 -{\bf k}\downarrow })$
where now we measure the momentum transfer relative to the interpocket momentum
${\bf k}+{\bf Q}_1$, {\it e.g.}  $c_{\beta_{1}{\bf k}+{\bf Q}_{1}\sigma}$.
Therefore, the Hamiltonian can be written as
\begin{eqnarray}
{\cal H} =
\sum\limits_{{\bf k} }
\hat{\psi}_{{\bf k} }^{\dagger}\hat{\beta}_{{\bf k}}
\hat{\psi}_{{\bf k} }
+
\sum\limits_{{\bf k}{\bf  k}^\prime }
\hat{\psi}_{{\bf k} }^{\dagger}\hat{U}_{ {\bf  k}{\bf  k}^\prime}
\hat{\psi}_{{\bf  k}^\prime}
\end{eqnarray}
where by defining $V^{i i}_{ {\bf  k}{\bf  k}^\prime}=\gamma u_0 $; $V^{i i^\prime}_{ {\bf  k}{\bf  k}^\prime}=\gamma u_Q $
and $J^{i i}_{zz}S_z=\gamma^\prime u_0 $; $J^{i i^\prime}_{zz}S_z=\gamma^\prime u_Q $, the matrices
$\hat{\beta}_{{\bf k}}$ and $\hat{U}_{ {\bf  k}{\bf  k}^\prime}$ are  defined as
\begin{eqnarray}
\hat{\beta}_{{\bf k}}=\left[
\begin{array}{cc}
\hat{\varepsilon}_{ {\bf  k}}^{\uparrow} & \hat{\Delta}_{\bf  k}
\\
\hat{\Delta}_{\bf  k} & \hat{\varepsilon}_{ {\bf  k}}^{\downarrow}
\end{array}
\right];
%
\hat{U}_{ {\bf  k}{\bf  k}^\prime}=
\left[
\begin{array}{cc}
\gamma^\prime+\gamma & 0
\\
0 & \gamma^\prime-\gamma
\end{array}
\right]
\otimes
\hat{I}_{ {\bf  k}{\bf  k}^\prime}.
\nonumber
\end{eqnarray}
$\otimes$ is the direct product of matrices and
\begin{widetext}
\begin{eqnarray}
 &&
\hat{\varepsilon}_{ {\bf  k}}^{\sigma} =\sigma\left[
\begin{array}{cccc}
\varepsilon^{\alpha_2}_{{\bf k}} & 0 & 0 & 0
\\
0 & \varepsilon^{\alpha_1}_{{\bf k}} & \sigma W & 0
\\
0 &  \sigma W &\varepsilon^{\beta_1}_{{\bf k}} & 0
\\
0 & 0 & 0 &\varepsilon^{\beta_2} _{{\bf k}}
\end{array}
\right];\;\;\;
%
\hat{\Delta}_{\bf  k}=\left[
\begin{array}{cccc}
\Delta^{\alpha_2}_{\bf  k} & 0  & 0 & 0
\\0 & \Delta^{\alpha_1}_{\bf  k}  & 0 & 0
\\0 & 0 & \Delta^{\beta_1}_{\bf  k}  & 0
\\0 & 0 & 0 & \Delta^{\beta_2}_{\bf  k}
\end{array}
\right];
\;\;\;
\hat{I}_{ {\bf  k}{\bf  k}^\prime}=\left[
\begin{array}{cccc}
u_0 & u_0 & u_{\bf Q} & u_{\bf Q}
\\u_0 & u_0 & u_{\bf Q}& u_{\bf Q}
\\u_{\bf Q} & u_{\bf Q} & u_0 & u_{\bf Q}
\\u_{\bf Q} & u_{\bf Q} & u_{\bf Q} & u_0
\end{array}
\right].
\nonumber
\end{eqnarray}
\end{widetext}
The spin index is $\sigma=\pm 1$  for spin up and down.
The overall $\sigma$ in front of the energy matrix appears due to the Nambu structure\cite{remark}.
Here, we assume that the intraband impurity scattering, $u_0$ differs from the
interband scattering between the bands separated by a large {\bf Q}, $u_{\bf Q}$.
The Green function (GF) matrix is obtained via
$G_{ {\bf  k}{\bf  k}^\prime}(\tau)=-\langle T \hat{\psi}_{{\bf k} }(\tau)
\hat{\psi}_{{\bf k}^\prime}^{\dagger}(0)\rangle$,
whence
\bea
G_{ {\bf  k}{\bf  k}^\prime}(\omega_n)
=G^{0}_{{\bf k}}(\omega_n)[\delta_{ {\bf  k}{\bf  k}^\prime} +
t_{ {\bf  k}{\bf  k}^\prime}(\omega_n)G^{0}_{{\bf k}^\prime}(\omega_n)],
\label{GF}
\eea
where $G^{0}_{{\bf k}}(\omega_n)=\left( i\omega_n-\hat{\beta}_{{\bf k}} \right)^{-1}$ is
the bare GF of the conduction electrons. Solving the Dyson equation for the T-matrix
\be t_{ {\bf  k}{\bf  k}^\prime}(\omega_n)=\hat{U}_{ {\bf  k}{\bf  k}^\prime}+\sum_{{\bf k}^{\prime \prime}} \hat{U}_{ {\bf  k}{\bf  k}^{\prime \prime}}G^{0}_{{\bf k}^{\prime\prime}}(\omega_n)
 t_{{\bf k}^{\prime \prime}{\bf k}^{\prime}}( \omega_n)
,\ee
the LDOS is obtained via analytic continuation
$i\omega_n\rightarrow \omega+i \sigma  0^{+}$ according to
\bea
N^{c}(\omega,{\bf r})&=&-\frac{1}{\pi} 
\mbox{Im} \, \mbox{Tr} \left[  G(r,r,\omega_n)\right]_{i\omega_n\rightarrow \omega+i \sigma 0^+}.
\label{DOS}
\eea
We recall that the interference between incident and scattered waves gives rise to the
spatial modulation of the amplitude of the total wave. This is then reflected in the LDOS.
We further set  $u_0 = 0.3t_{\alpha}$.

\section{Results}
In what follows we discuss systematically the different ordered phases of the iron pnictides as shown in the phase diagram of Fig.(\ref{Phase}) and present our calculated results for the total spectral function [Eqs.(\ref{GF})--(\ref{DOS})].
The spectral function $\sum_{\sigma} \mbox{Im}\, \mbox{Tr}\,
G_{0\sigma}({\bf k},\omega)$ of the clean system  is always shown on the left panel. This allows to
trace easily the most important scattering vectors that appear in the QPI corrections $\sum_{{\bf k}\sigma} \mbox{Im}\, \mbox{Tr} \left[G_{0\sigma}({\bf k},\omega)
t_{\sigma}({\bf k}, {\bf k+q},\omega) G_{0\sigma}({\bf k+q},\omega)\right]$.
We discuss the QPI maps
for the case of non-magnetic and magnetic impurities and show the difference between them.
The latter is particular important for identifying the symmetry of the superconducting gap.

\subsection{Spin-Density wave phase}

\begin{figure}
 \centering
\includegraphics[angle=0,width=\linewidth]{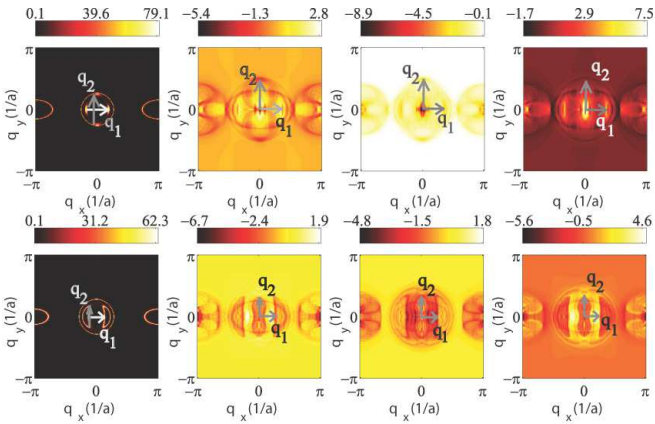}
\caption{(color online) Constant energy intensity maps at $\omega=20 meV$ (first row) and  $w=-20 meV$ (second row) of the spectral density, $\sum_{\sigma} \mbox{Im}\, \mbox{Tr}\,
G_{0\sigma}({\bf k},\omega)$, (first panel from the left) and
QPI, $\sum_{{\bf k}\sigma} \mbox{Im}\, \mbox{Tr} \left[G_{0\sigma}({\bf k},\omega)
 t_{\sigma}({\bf k}, {\bf k+q},\omega) G_{0\sigma}({\bf k+q},\omega)\right]$
for non-magnetic (second panel) and magnetic (third panel) impurities obtained as described in the text.
The difference between QPI for non-magnetic and magnetic  impurities is displayed in the right panel.
The color bars refer to the intensity in units of states/eV.
The SDW order parameter is fixed at $W=40 meV$ for zero doping.}
\label{SDW_d0}
\end{figure}

 \begin{figure}[h]
\includegraphics[angle=0,width=\linewidth]{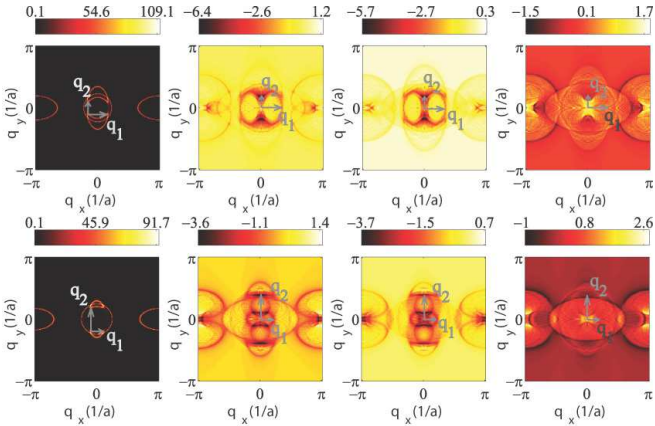}
\caption{(color online) Constant energy intensity maps at $\omega=20 meV$ (first row) and  $w=-20 meV$ (second row) of the spectral density  and
QPI,
for non-magnetic (second panel) and magnetic (third panel) impurities obtained as described in the text.
The difference between QPI for non-magnetic and magnetic  impurities is displayed in the right panel.
For  SDW (region I in Fig.\ref{Phase}) with $W = 40 meV$ and with  $7\%$  e-doping,  for $\omega=20 meV$ (first row) and  $w=-20 meV$ (second row). The color bars refer to the intensity in units of states/eV.
The SDW order parameter is fixed at $W=40 meV$ for $7\%$ electron doping. }
\label{SDW_d7}
\end{figure}

First, we review our results for the QPI in the SDW phase and its changes due
to doping  [regime (I) in Fig.(\ref{Phase})] by showing in Figs.(\ref{SDW_d0}) and Figs.(\ref{SDW_d7})
the QPI for zero and $7\%$ electron doping, respectively.
A consequence of SDW order in parent
iron-based superconductors is that the C$_4$ symmetry of the normal state is continuously broken.
The resulting Fermi surface consists of small pockets  which originate due to mixing of one of
the hole pockets centered around the $\Gamma-$point and the elliptic electron pocket centered around the
$(0,\pi)$ point of the BZ as shown in the left panel. The scattering between the resulting boomerang-like
structures (the most characteristic wave vectors {\bf q}$_1$ and ${\bf q}_2$ are indicated by arrows)
dominates the QPI at low energies and occurs for both non-magnetic and magnetic
impurity scattering\cite{Knolle2010,chuang}.

Note that this reduction from $C_4$ to C$_2$ symmetry in the magnetically ordered state was originally
interpreted as a sign of the underlying quasi-one-dimensional electronic structure and the electronic nematic order\cite{chuang,wang_new} .
In our case, however, it appears just as a result of the translational symmetry breaking induced by the ($0,\pi$) antiferromagnetic state on the
two-dimensional electronic structure. We also find that the $C_4$ symmetry is restored
for bias energies exceeding twice the SDW gap value. Recently this prediction was confirmed experimentally\cite{Li10} again signifying the dominant role of the magnetic order in breaking the $C_4$ symmetry.

For large enough $W$ the FSs of the bands involved
in the SDW are completely gapped and the same is true for the spectral densities at low energies.
In this case, QPI will be  determined by the bands which are not involved in the SDW, and, therefore,
its structure will not show strong quasi--1D character. This may explain why the pure AF SDW state does
not show any $C_2$-symmetric structure in the parent compounds where the magnetic moment (and the
corresponding SDW gap) is quite large. Only when it is reduced upon doping, the bands involved in
the SDW are located close to the FS so that the  QPI structure, described above, becomes visible. Another apparent
effect is the particle-hole asymmetry of the QPI which is the result of a rotation of the SDW induced
pockets from positive to negative energies by $90^{\circ}$ degrees.
The asymmetry is $90^{o}$ rotated in the case of electron doping (see Fig.\ref{SDW_d7}).  Due to the larger size of the electron
pockets at $(\pm\pi,0)$ the SDW induced small pockets are pronounced and dominate the scattering
features in the QPI maps. For low negative energies the electron pocket is still present and the
90$^{\circ}$  rotation of the one dimensional features shifts to lower energies.

\subsection{Superconducting phase}
 \begin{figure}[h]
\includegraphics[angle=0,width=\linewidth]{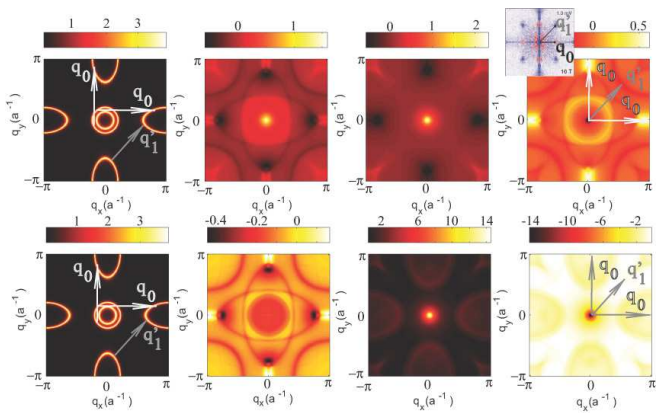}
\caption{(color online) The spectral density and the QPI maps for non-magnetic and magnetic impurity scattering
(from left to right) at 20meV
in  SC state with nodeless $s^{+-}$-wave (upper panel) and isotropic $s^{++}$-wave (lower panel) symmetries of the
superconducting gap.  We set $\Delta=20 meV$, $13\%$  e-doping.
The difference between QPI for non-magnetic and magnetic  impurities is displayed on the right panel. The arrows show
scattering between electron and hole bands at {\bf q}$_0 = (\pi, 0)$, $(0,\pi)$. The inset in the upper panel shows the
the experimental data taken from Ref.\onlinecite{hanaguri2010}.
}
\label{fig:sc}
\end{figure}

Next, we discuss the QPI maps in the superconducting state.
As mentioned above,  the intensity of the QPI is affected by the relative sign of the order parameter between the
FSs involved in the scattering via the Bogolyubov coherence factors representing the formation of Cooper pairs and
new quasiparticles, which are coherent superpositions of electrons and holes.
Coherence factors characterize how the scattering of a superconducting quasiparticle off a given scatterer differs
from the scattering of a bare electron off the same scatterer. Generally, the momentum-dependent order parameter
enters into the expression for the coherence factor so that studies of scatterings of quasiparticles with different
momenta can provide important information on the phase of the superconducting order parameter in momentum space.
Originally, this idea was put forward for the cuprates\cite{coleman,hanaguri_cup,franz}.
In particular, for potential (non-magnetic) scattering at wave vector ${\bf q}$
the corresponding coherence factor is smaller for those {\bf q} which connect
parts of the FS with the same sign of the SC gap.

Therefore, the QPI intensity appears
only for sign-reversing momenta. In iron-based superconductors with
$s^{+-}$-wave symmetry of the superconducting order parameter, this occurs for
$(\pi,0)$ and $(0,\pi)$ momenta. Then
for magnetic (time-reversal) scattering, the QPI is negative for the wave vectors
$(\pi,0)$ and $(0,\pi)$ and
positive for momenta $(\pm \pi,\pm \pi)$ as the latter connects FSs with the same signs of the order parameter.
Overall, the effect of the sign change of the superconducting gap can be most efficiently seen if one plots the difference of QPI for non-magnetic and magnetic impurities.

As a result, a comparison of the QPI maps for magnetic and non-magnetic scattering
yields important information on the symmetry of the superconducting order parameter. In practice, the most
straightforward way to perform this comparison in a type II superconductor is to apply an external magnetic field where the resulting vortices act in part
as magnetic scatterers\cite{hanaguri2010,hanaguri_cup}. Thus, by comparing the results with and without magnetic field the symmetry
of the superconducting order can be deduced.

In Fig.\ref{fig:sc} we show the QPI maps for the nodeless $s^{+-}$-wave ($\Delta'=$0) and isotropic $s^{++}$-wave symmetry.
Clearly by subtracting the QPI for the non-magnetic and magnetic scattering shown in the right panel, the difference
between the two gaps becomes apparent with regard to the interband scattering (i) at ${\bf q}_0 = (\pm \pi,0)$ or
($0,\pm \pi$) which occurs for the scattering between electron and hole bands and (ii) diagonal
${\bf q}^{\prime}_1$-scattering between two electron pockets.
As the sign of the superconducting gap is opposite between the $\alpha$ and the $\beta$-bands the difference in the
QPI at {\bf q}$_0$  between non-magnetic and magnetically induced scattering is positive. By contrast, that same
difference in the scattering at
${\bf q}_1$ is negative as the order parameter on the electron pockets possesses the same sign. Both features  are
consistent with the QPI measured in the superconducting state of Fe(Se,Te) and Ba(Fe$_{1-x}$Co$_x$)$_2$As$_2$
compounds\cite{hanaguri2010,teague}. In particular, in the inset of Fig.\ref{fig:sc} we show the SI-STM  data taken
at 10T in FeTe$_{0.4}$Se$_{0.6}$ compound. Observe the difference in the intensities in the external magnetic field of
10T for the scattering at
{\bf q}$_0$ and ${\bf q}^{\prime}_1$ which is consistent with our calculations. Note also
that in the difference plot the details of the Fermi surface scattering
are very weakly visible, including the scattering between the electron pockets. Instead the largest effect occurs due the sign change of the gap and
the corresponding coherence factors.  For comparison, in Fig.\ref{fig:sc} (lower panel) we show the
same calculations for
$s^{++}$-wave order parameter. The striking distinction with the $s^{+-}$-wave case is  that
the difference between non-magnetic and magnetic impurity induced QPI is, as expected,
featureless.

 \begin{figure}[h]
\includegraphics[angle=0,width=0.5\linewidth]{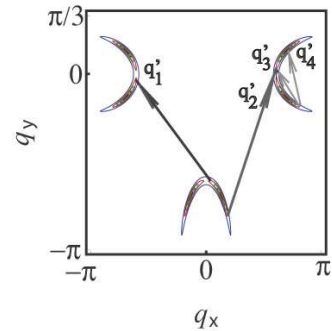}
\caption{(color online) Calculated spectral density in the superconducting state with nodal $s^{+-}$-wave symmetry
for $\Delta^{\prime} = 3\Delta$. The arrows refer to some intra and inter electron pocket scattering wavevectors
({\it i.e.}, wavevectors joining regions with large DOS). Observe
that ${\bf q}^{\prime}_1$, and ${\bf q}^{\prime}_2$, ${\bf q}^{\prime}_3$, ${\bf q}^{\prime}_4$ are the scattering
wave vectors for the same and the opposite
signs of the superconducting order parameter, respectively. The evolution of the banana-shape structures is shown
for +5
(black curve), +10 (brown curve), +15 (green curve), +20 (red curve), and +30 (blue curve) meV.}
\label{fig5}
\end{figure}
In the next step we show the effect of higher harmonics in the $s^{+-}$-wave function.
By taking a non-zero $\Delta^{\prime}$, gap minima or even nodes form on the electron pockets.
In this regard the evolution of the spectral density with energy will resemble somewhat the structure found in
nodal $d$-wave superconductors. In particular, due to the nodal structure of the superconducting  gap,
the spectral density with increasing energy will show a banana-shaped structure with new
scattering momenta associated with the large density of states at the edges of the banana as shown in Fig.\ref{fig5}.
%
 \begin{figure}[h]
\includegraphics[angle=0,width=\linewidth]{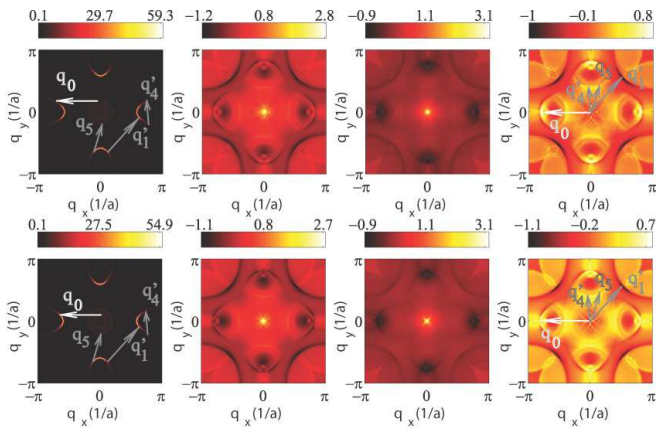}
\caption{(color online) (color online) The spectral density and the QPI maps for non-magnetic and magnetic impurity scattering
(from left to right) at +20meV (upper panel) and +10meV (lower panel)
in  SC state with nodal $s^{+-}$-wave order parameter of the
superconducting gap ($\Delta^{\prime}=3\Delta$) and $13\%$ of the electron doping.
The difference between QPI for non-magnetic and magnetic  impurities is displayed on the right panel.}
\label{fig6}
\end{figure}
We selectively show only some of the scattering wave vectors on this plot which we later identify on the QPI maps.
Note that ${\bf q}^{\prime}_1$ refers to the scattering with the same sign of the order parameter while other
wavevectors, ${\bf q}^{\prime}_2$, ${\bf q}^{\prime}_3$, and ${\bf q}^{\prime}_4$ connect regions with
the superconducting order parameter of opposite sign. In principle these additional features
should be most pronounced in the
difference of QPI intensities for magnetic and non-magnetic scattering.

The results for the spectral density and QPI maps are shown in Fig.\ref{fig6}. Due to relatively large
quasiparticle lifetime used in the calculations ($\sim 4$meV) the banana-shaped structures are not well
resolved and overall the resulting QPI
does not have well-defined momenta as those shown in Fig.\ref{fig6}. However, the anisotropy of the superconducting
gap is clearly observed in the difference between magnetic and non-magnetic impurity induced QPI, as shown on the right
panel. First, we again find the scattering along the diagonal momentum at {\bf q}$^{\prime}_1$ which was also present
in the case of a nodeless $s^{+-}$-wave symmetry. It is, however, additionally enhanced due to
banana-shape structure and the corresponding high density of states at the edges of these bananas.
Similarly, the interband induced scattering between the $\alpha-$ and the $\beta$-bands (indicated by {\bf q}$_0$)
acquires more structure. In addition, the maximum of intensity
shifts from $(\pm \pi, 0)$ and $(0, \pm \pi)$ to the smaller momenta due to the higher harmonics
in the $s^{+-}$-wave gap. Additional peaks at small momenta include the
scattering at ${\bf q}^{\prime}_4$ within electron pockets and also the scattering between electron and hole pockets
connecting states with opposite sign of the superconducting gap shown by ${\bf q}_5$.
Overall, the structure of QPI becomes considerably richer, which should
allow identification of the nodal structure of $s^{+-}$-wave superconducting order. One has to bear in mind, however,
that the higher harmonics in the $s^{+-}$-wave symmetry with corresponding scattering at {\bf q}$^{\prime}_4$ and
{\bf q}$_5$ should appear at larger doping. Note, to prove unambiguously whether these originate from the nodal $s^{+-}$-wave
order parameter measurements in an external magnetic field are required.

\subsection{Coexistence phase with SDW and $s^{+-}$ SC order}
 \begin{figure}[h]
\includegraphics[angle=0,width=\linewidth]{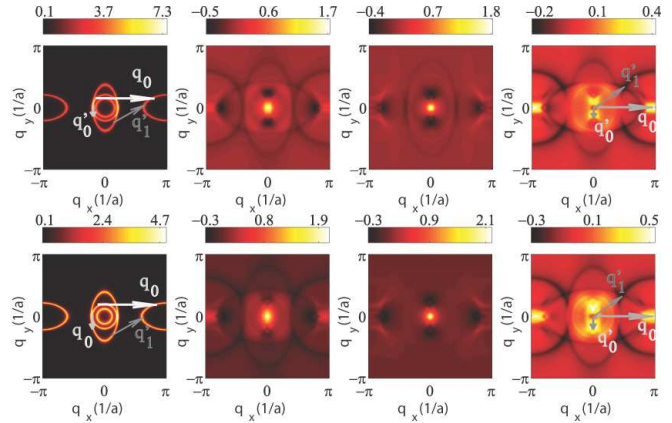}
\caption{(color online) Calculated spectral density and the QPI maps at $\omega = +20$meV in
the coexistence regime of SDW + SC. The upper panel refers to the regime II of the phase diagram (Fig.\ref{Phase}) with
$W = 40 meV$, $\Delta=20 meV$
and $10\%$  e-doping. The lower panel describes the regime III with $W = 10 meV$, $\Delta=20 meV$
and $12\%$  e-doping.}
\label{fig7}
\end{figure}

Finally, we consider the coexistence regime with both microscopic SDW and
$s^{+-}$-wave superconducting order. In  Fig.\ref{fig7}  we examine two different situations which correspond
to regimes (II) and (III) of the phase diagram shown in Fig.\ref{Phase}, respectively. In regime (II) the
SDW order parameter is larger than the SC one and we set $W=2\Delta$. In regime (III), which corresponds to higher
doping, the situation reverses and we use $W=\Delta/2$. For the sake of
simplicity and also because it is expected that the nodal $s^{+-}$-wave order occurs only at higher
doping\cite{chubukov_nodal,thomale}
we neglect the effect of the higher harmonics in the gap, {i.e.} we put $\Delta^{\prime} = 0$.

For the case of a larger SDW gap, the scattering has a pronounced $C_2$ symmetry and
the only effect of superconductivity shown in the QPI,
as compared to the pure  SDW case, is to
reduce  the overall intensity of the scattering peaks (note the reduced intensity on the color bars).
The latter occurs due to additional gapping of the bands.
The structures visible in the upper panel of
Fig.\ref{fig7} are similar to those seen Fig.\ref{SDW_d0}, with some extra smearing due to the superconducting gap.
There are, however, additional structures which become particularly visible
in the difference between QPI maps for magnetic and non-magnetic scattering,
which are signatures of the nodeless $s^{+-}$-wave gap. Note that
these features strengthen in the case when the SC gap is larger than the SDW one, as is seen in the lower panel
of Fig.\ref{fig7}.

The origin of these additional structures can be traced back to the effect of the SC $s^{+-}$ order
which is superimposed on the SDW order. First, notice that in the difference plot there is always
an enhanced intensity at ${\bf q}_0 = (\pm \pi,0)$ momenta. This is due to the fact that the electron and the
hole band not involved in the SDW state with $(0,\pi)$ order
have a different sign of the superconducting gap. Then for the very same reason as
in the pure SC state with nodeless $s^{+-}$-wave order parameter,  the difference between the magnetic and non-magnetic
impurity-induced QPI shows the strongest feature for interband scattering at $(\pm \pi,0)$ momenta. At the same
time, new features arise due to the fact that one of the electron
and one of the hole bands are mixed in the SDW state with $(0,\pi)$ ordering. However, both of these bands
still possess an opposite sign of the superconducting gap, thus leading to a structure in the difference plot
at small momenta. Remember that
in the pure SC state with s$^{+-}$-wave order, these peaks would occur for $(0,\pm \pi)$ momentum but now
this is a reciprocal wave vector of the SDW state. As a result, this scattering is
'translated' to that
around $q \sim 0$ momentum. This clearly shows that the regime of microscopic coexistence of SDW+SC
can be nicely observed in the FS-STM data. The same occurs for the scattering between two electron ($\beta$)-bands.
Originally along the diagonal of the BZ ({\bf q}$^{\prime}_1$ in the nodeless $s^{+-}$-wave symmetry),
it is now shifted to a different momentum due to the folding of the electron band involved in the SDW formation.


\section{Conclusion}
In this paper we have systematically calculated quasiparticle interference effects due to magnetic and
nonmagnetic impurities for the whole phase diagram of the iron pnictide superconductors. We have shown
that in the SDW phase the QPI shows quasi one-dimensional features as measured in recent experiments\cite{chuang}.
The $C_4$ symmetry is restored for energies
larger than twice the SDW gap value.
Furthermore, analyzing the superconducting state we have  shown
that the nodeless as well as the nodal $s^{+-}$-wave symmetry of the superconducting gap can be clearly
identified in SI-STM experiments and distinguished from the isotropic $s$-wave gap. In particular,
the scattering between the electron and the hole pockets at {\bf q}$_0 = (\pm \pi,0)$ [$(0,\pm \pi)$]
as well as the scattering between the electron pockets at {\bf q}$^{\prime}_1$ becomes quite pronounced in
the difference of the QPI maps between magnetic and non-magnetic impurities, a feature found also in the
experimental data\cite{hanaguri2010}.  We have
demonstrated that a non-trivial angular dependence of the $s^{+-}$-wave gap induced by higher harmonics results
in banana-shape structures in the QPI maps not present in the simplest version of $s^{+-}$-wave symmetry.
A large density of states associated with the edges of these bananas is responsible for an additional
low ${\bf q}$-scattering along the bond direction as well diagonal ${\bf q}$-scattering
patterns in the QPI. Again these scatterings are especially pronounced in the difference
plots between the magnetic and non-magnetic impurity scattering. In the regime of coexistence of
SDW and SC with $s^{+-}$-wave
order our study has revealed additional characteristic features of the QPI which may help to identify this
phase in iron-based superconductors. It includes: (a) the $C_2$ anisotropy of the QPI maps in the coexistence regime
and (b) the low ${\bf q}$-interband scattering for the bands with opposite sign of the $s^{+-}$
superconducting gap.

We acknowledge useful discussions with A.V. Chubukov, T. Hanaguri, H. Takagi and are thankful to
I.I. Mazin for the critical remarks. I.E. acknowledges the support from the
RMES Program [Contract No. 2.1.1/2985].


\begin{thebibliography}{99}
\bibitem{remark} Note, this definition of the Nambu spinor differs from the one used for the
pure SDW state\cite{Knolle2010} where $
\hat{\psi}_{{\bf k}}^{\dagger}=
(c_{\alpha_2 {\bf k}\uparrow }^{\dagger},c_{\alpha_1 {\bf k}\uparrow }^{\dagger},c_{\beta_1{\bf k}\uparrow  }^{\dagger},c_{\beta_2{\bf k}\uparrow  }^{\dagger}, c_{\alpha_2 {\bf k}\downarrow }^{\dagger},c_{\alpha_1 {\bf k}\downarrow }^{\dagger}, c_{\beta_1{\bf k}\downarrow  }^{\dagger},c_{\beta_2{\bf k}\downarrow  }^{\dagger})$. This yields an opposite structure of the $\hat{\beta}_{{\bf k}}$ for the magnetic and non-magnetic impurity.
\end{thebibliography}
\end{document}